\documentstyle[preprint,aps]{revtex}
\begin{document}
\draft
\preprint{hep-ph/0106074}

\title{Detailed Balance and Sea-Quark Flavor Asymmetry of Proton}

\author{Yong-Jun Zhang\footnote{zyj@pubms.pku.edu.cn}}
\address{Department of Physics, Peking University, Beijing 100871, China}

\author{Bin Zhang}
\address{Department of Physics, Tsinghua University, Beijing 100084, China}

\author{Bo-Qiang Ma\footnote{mabq@phy.pku.edu.cn}}
\address{CCAST (World Laboratory), P.O.~Box 8730, Beijing
100080, China \\
Department of Physics, Peking University, Beijing 100871, China
\footnote{Mailing address}}


\maketitle
\begin{abstract}
In this study, the proton is taken as an ensemble of quark-gluon
Fock states. Using the principle of detailed balance, the
probabilities of finding every Fock states of the proton are
obtained without any parameter. A new origin of the light flavor
sea quark asymmetry, i.e., $\bar{u} \not= \bar{d}$, is given as a
pure statistical effect. It is found that $\bar{d}-\bar{u} \approx
0.124$, which is in surprisingly agreement with the experimental
observation.

\end{abstract}

\vfill


\narrowtext

The composition of hadrons in terms of the fundamental quark and gluon degrees
of freedom is a central focus of hadronic physics. In the light-cone
Fock state description of bound states \cite{LC},
each physical hadron state is expended
by a complete set of quark-gluon Fock states as
\begin{equation}
\left| h \right>=\sum_{i,j,k} c_{i,j,k}\left| \{q\}, \{i,j,k\}, \{l\} \right>,
\label{FockStates}
\end{equation}
where $\{q\}$ represents the valence quarks of the hadron, $i$ is
the number of quark-antiquark $u\bar{u}$ pairs, $j$ is the number
of quark-antiquark $d \bar{d}$ pairs, $k$ is the number of gluons,
and $\{l\}$ represents the heavy flavor ($s$, $c$, $b$, and $t$)
$q \bar{q}$ pairs which will be neglected at first in this study.
There have been many progress in understanding the hadron
structure in terms of the underlying quark and gluon degrees of
freedom. For example, we can obtain the lowest valence Fock state
$\left| \{q \}\right>$ by the quark model \cite{QM}. We can also
model the next higher $\left| \{q \},\{q\bar{q}\}\right>$ Fock
states by the energetically-favored baryon-meson fluctuations
\cite{Bro96}. The Fock states with gluons, $\left| \{q
\},\{g\}\right>$, have been also studied \cite{InGluon}. However,
we are still lacking a comprehensive understanding concerning the
completed set of Fock states for hadrons, even for the proton.

In this work, we will provide a complete set of Fock states
for the proton from a simple statistical consideration. By (\ref{FockStates}),
it means that we take the proton as an ensemble of quark-gluon
Fock states. The probability to find the proton in the
$\left|\{q\},\{i,j,k\}\right>$ Fock state is
\begin{equation}
\rho_{i,j,k}=\left| c_{i,j,k}\right|^2,
\end{equation}
where $\rho_{i,j,k}$ satisfies the normalization condition,
\begin{equation}
\sum_{i,j,k} \rho_{i,j,k}=1. \label{unit}
\end{equation}
Therefore the parton numbers of quarks and gluons in the proton are
\begin{equation}
\begin{array}{ll}
u=u_{v}+\sum_{i,j,k} i\, \rho_{i,j,k},\\
d=d_{v}+\sum_{i,j,k} j \, \rho_{i,j,k},\\
\bar{u}=\sum_{i,j,k} i\, \rho_{i,j,k},\\
\bar{d}=\sum_{i,j,k} j\, \rho_{i,j,k}, \\
g=\sum_{i,j,k} k\, \rho_{i,j,k},\\
\end{array}
\end{equation}
where $u_{v}=2$ and $d_{v}=1$ are the valence quark numbers of the proton.
These parton numbers can be measured by deep inelastic scattering of leptons
on the proton target. The quarks and gluons in the Fock states are
the ``intrinsic" partons of the proton, since they are multi-connected non-perturbatively
to the
valence quarks \cite{Bro96}. Such partons are
different from the ``extrinsic"
partons generated from
the QCD hard bremsstrahlung and gluon-splitting
as part of the lepton scattering interaction.

Using statistical property of the ensemble, the probability
$\rho_{i,j,k}$ can be calculated without any parameter in the
following way. Let us trace one system (a proton) in the big
ensemble (a set of quark-gluon Fock states). At the time $t=0$,
the system is in one state of a subensemble
($A=\left|\{q\},\{i,j,k\}\right>$). After a while, at the time
$t=\delta t$, the state of this system has three probabilities: (1) does
not change; (2) changes to another state that belongs to the same
subensemble ($A$); (3) changes to a state that belongs to a different
subensemble ($B=\left|\{q\},\{i',j',k'\}\right>$). In the last
case, the density of subensemble ($A$) decreases ($\rho_A
\downarrow$). On the other hand, some states in subensemble ($B$)
will be changed to subensemble ($A$) during the same period, and
increases the density of subensemble ($A$) ($\rho_A \uparrow$).
These two processes, which are in opposite direction for each
other, just compensate for each other, so $\rho_A$ keeps unchanged.
This is the so called detailed balance, which demands every two
subensembles to balance with each other. So we have
\begin{equation}
\rho_{i,j,k}\left|\{q\},\{i,j,k\}\right>\stackrel{\rm
balance}{\Longleftrightarrow}
\rho_{i^\prime,j^\prime,k^\prime}\left|\{q\},\{i^\prime,j^\prime,k^\prime\}\right>,
\end{equation}
\begin{equation} 
\rho_{i,j,k}N\left\{\left|\{q\},\{i,j,k\}\right>\rightarrow
\left|\{q\},\{i^\prime,j^\prime,k^\prime\}\right>\right\}
\equiv
\rho_{i^\prime,j^\prime,k^\prime}N\left\{\left|\{q\},\{i^\prime,j^\prime,k^\prime\}
\right>\rightarrow
\left|\{q\},\{i,j,k\}\right>\right\},
\end{equation}
\begin{eqnarray} 
\frac{\rho_{i,j,k}}{\rho_{i^\prime,j^\prime,k^\prime}}
\equiv
\frac{N\left\{\left|\{q\},\{i^\prime,j^\prime,k^\prime\}
\right>\rightarrow
\left|\{q\},\{i,j,k\}\right>\right\}}
{N\left\{\left|\{q\},\{i,j,k\}\right>\rightarrow
\left|\{q\},\{i^\prime,j^\prime,k^\prime\}\right>\right\}},
\label{ratio}
\end{eqnarray}
where $N\{A \rightarrow B\}$ implies the number of processes that transfer $A$ into
$B$, i.e., the transfer rate of $A \to B$.

In order to know $\rho_{i,j,k}$, the probability of finding the
proton in the Fock state $A$, all what we need to know is the
ratio between the transfer rate of $A \to B$ and the transfer rate
of $B \to A$. In the realistic situation,  the transfer rate is
very complicated and it involves many factors, such as: the proton
size, the quark mass, the parton numbers, the quantum numbers
(color,spin,flavor),  cross sections of sub-processes, exchange
symmetry (Pauli blocking), the velocity and even the details of
the quantum wave functions of partons. Here, among all these
factors, we only take into account the number of partons from a
pure statistical consideration. We also neglect the interaction
involving $g\Leftrightarrow gg$ at first. We will find,
surprisingly, that our simplified calculation gives the correct
light flavor sea-quark asymmetry, i.e., $\bar{u} \not= \bar{d}$,
which has been observed in experiments
\cite{FlavorAsymmetry,NMC91,NA51,HERMES,E866a,E866}. This implies
that the number of valence quarks controls the statistical property
of the proton.

The transfer between two subensembles has two ways: splitting and
recombination. Splitting rate is proportional to the number of
partons that may split. Recombination rate is proportional to both
the number of those two kinds of partons that may recombine. So we
have the following two kinds of relations:

\noindent
(1) Detailed balance involving $q\Leftrightarrow qg$ gives the
relation
\begin{eqnarray}
|uud\rangle{\footnotesize\begin{array}{c}3\\
{\Large\rightleftharpoons}
\\ 1\times3 \end{array}}|uudg\rangle,\\
|uudg\rangle{\footnotesize\begin{array}{c}3\\
{\Large\rightleftharpoons}
\\ 2\times3 \end{array}}|uudgg\rangle ,\\
|uudgg\rangle{\footnotesize\begin{array}{c}3\\
{\Large\rightleftharpoons}
\\  3\times3 \end{array}}|uudggg\rangle,\\
\cdots\\ |uud\bar{u}u\rangle{\footnotesize\begin{array}{c}5\\
{\Large\rightleftharpoons}
\\  1\times5 \end{array}}|uud\bar{u}ug\rangle,\\
\cdots \\ |\{q\},\{i,j,k-1\}\rangle{\footnotesize\begin{array}{c}3+2i+2j\\
{\Large\rightleftharpoons}
\\ (3+2i+2j)k \end{array}}|\{q\},\{i,j,k\}\rangle.
\end{eqnarray}
then a general formula can be derived using formula
(\ref{ratio}),
\begin{eqnarray}
\frac{\rho_{i,j,k}}{\rho_{i,j,k-1}}=\frac{1}{k},\\
\frac{\rho_{i,j,k}}{\rho_{i,j,0}}=\frac{1}{k!}\label{number
gluon}.
\end{eqnarray}

\noindent
(2) Detailed balance involving $g\Leftrightarrow \bar{q}q$ gives
the relation
\begin{eqnarray}
|uudg\rangle{\footnotesize\begin{array}{c}1\\
{\Large\rightleftharpoons}
\\ 1\times3 \end{array}}|uud\bar{u}u\rangle,\\
|uudg\rangle{\footnotesize\begin{array}{c}1\\
{\Large\rightleftharpoons}
\\ 1\times2 \end{array}}|uud\bar{d}d\rangle,\\
|uud\bar{u}ug\rangle{\footnotesize\begin{array}{c}1\\
{\Large\rightleftharpoons}
\\ 2\times4 \end{array}}|uud\bar{u}u\bar{u}u\rangle,\\
|uud\bar{u}ug\rangle{\footnotesize\begin{array}{c}1\\
{\Large\rightleftharpoons}
\\ 1\times2 \end{array}}|uud\bar{u}u\bar{d}d\rangle,\\
\cdots \\ |\{q\},\{i-1,j,1\}\rangle{\footnotesize\begin{array}{c}1\\
{\Large\rightleftharpoons}
\\ i(i+2) \end{array}}|\{q\},\{i,j,0\}\rangle, \\
|\{q\},\{i,j-1,1\}\rangle{\footnotesize\begin{array}{c}1\\
{\Large\rightleftharpoons}
\\ j(j+1) \end{array}}|\{q\},\{i,j,0\}\rangle.
\end{eqnarray}
Then a general formula can be derived using formula
(\ref{ratio}),
\begin{eqnarray}
\frac{\rho_{i,j,0}}{\rho_{i-1,j,1}}=\frac{1}{i(i+2)},\\
\frac{\rho_{i,j,0}}{\rho_{i,j-1,1}}=\frac{1}{j(j+1)}.
\end{eqnarray}
Using relation $\rho_{i,j,1}=\rho_{i,j,0}$ gotten from formula
(\ref{number gluon}) to these two new formulae, we get
\begin{eqnarray}
\frac{\rho_{i,j,0}}{\rho_{0,j,0}}=\frac{2}{i!(i+2)!},\\
\frac{\rho_{i,j,0}}{\rho_{i,0,0}}=\frac{1}{j!(j+1)!},\\
\frac{\rho_{i,j,0}}{\rho_{0,0,0}}=\frac{2}{i!(i+2)!j!(j+1)!}\label{quark}.
\end{eqnarray}
Combing formulae (\ref{number gluon}) and (\ref{quark}), we obtain the
general formula
\begin{equation}
\frac{\rho_{i,j,k}}{\rho_{0,0,0}}=\frac{2}{i!(i+2)!j!(j+1)!k!}\label{aijk}.
\end{equation}
Then from the normalization condition (\ref{unit}) and
formula (\ref{aijk}), all $\rho_{i,j,k}$ can be calculated as
shown in TABLE \ref{table1}.

 From TABLE \ref{table1}, we get the numbers of intrinsic
gluons and sea quarks of the proton,
\begin{eqnarray}
\bar{u}=\sum_{i,j,k}
i\rho_{i,j,k}=0.308,\\ \bar{d}=\sum_{i,j,k} j\rho_{i,j,k}=0.432,\\
 g=\sum_{i,j,k} k\rho_{i,j,k}=0.997,\\
 \bar{d}-\bar{u}=0.124.
 \end{eqnarray}
These parton numbers should be considered to work at a scale for
the ``intrinsic" partons, and the details need further studies.
It is very interesting that only from the pure statistical consideration,
we get an asymmetry between the light flavor $u$ and $d$ sea quarks, i.e.,
$\bar{u} \not= \bar{d}$, which has been observed in deep inelastic
scattering and Drell-Yan experiments \cite{FlavorAsymmetry,NMC91,NA51,HERMES,E866a,E866}.
Moreover, the
flavor sea-quark asymmetry $\bar{d}-\bar{u}$ can be checked
by experiments directly because its $Q^2$ dependence is small. It is a surprise
that our
result is in excellent agreement with the recent experimental result \cite{E866}
$\bar{d}-\bar{u}=0.118 \pm 0.012$. This good agreement indicates that
the principle of detailed balance  plays an essential role in the structure of
proton.

We now check whether the inclusion of the process
$g\Leftrightarrow gg$ can change the above conclusion. We make the similar analysis
on the two kinds of relations:

\noindent
(1) Detailed balance involving $q\Leftrightarrow
qg$ and $g\Leftrightarrow gg$ gives the relation
\begin{eqnarray}
|uud\rangle{\footnotesize\begin{array}{c}3\\
{\Large\rightleftharpoons}
\\ 1\times3 \end{array}}|uudg\rangle,\\
|uudg\rangle{\footnotesize\begin{array}{c}3+1\\
{\Large\rightleftharpoons}
\\ 2\times3+1 \end{array}}|uudgg\rangle ,\\
|uudgg\rangle{\footnotesize\begin{array}{c}3+2\\
{\Large\rightleftharpoons}
\\  3\times3+3 \end{array}}|uudggg\rangle,\\
\cdots\\ |uud\bar{u}u\rangle{\footnotesize\begin{array}{c}5\\
{\Large\rightleftharpoons}
\\  1\times5 \end{array}}|uud\bar{u}ug\rangle,\\
\cdots \\ |\{q\},\{i,j,k-1\}\rangle{\footnotesize\begin{array}{c}3+2i+2j+k-1\\
{\Large\rightleftharpoons}
\\ (3+2i+2j)k+C_k^2 \end{array}}|\{q\},\{i,j,k\}\rangle,
\end{eqnarray}
where $C_k^2=\frac{k(k-1)}{2}$. Then a general formula can be
derived using formula (\ref{ratio}),
\begin{equation}
\frac{\rho_{i,j,k}}{\rho_{i,j,k-1}}=\frac{3+2i+2j+k-1}{(3+2i+2j)k+\frac{k(k-1)}{2}}\label{g-gg}.
\end{equation}

\noindent
(2) Detailed balance involving $g \Leftrightarrow \bar{q}q$ has
no effect by the process $g\Leftrightarrow gg$ so the
Eq.(\ref{quark}) remains unchanged.

Combining (\ref{g-gg}), (\ref{quark}), and the normalization
condition (\ref{unit}), all $\rho_{i,j,k}$ can be calculated as
shown in TABLE \ref{table2}. From TABLE \ref{table2}, we get the numbers of intrinsic
gluons and sea quarks of the proton,
\begin{eqnarray}
\bar{u}=\sum_{i,j,k}
i\rho_{i,j,k}=0.304,\\ \bar{d}=\sum_{i,j,k} j\rho_{i,j,k}=0.426,\\
 g=\sum_{i,j,k} k\rho_{i,j,k}=1.109,\\
 \bar{d}-\bar{u}=0.123,
 \end{eqnarray}
which are very close to the situation without considering  $g \Leftrightarrow g g$,
except for the gluon number with a bigger difference of 10\%.
Here the flavor sea-quark asymmetry $\bar{d}-\bar{u}$ is also in good agreement
with the recent experimental result
$\bar{d}-\bar{u}=0.118 \pm 0.012$.

We point out here that the heavy quark-antiquark pairs are not considered in this
paper.
Taking them into account is not so simple and may introduce some parameters, and
this is beyond the simple spirt of this paper.
So we neglect them in this study,
though there have been suggestions of ``intrinsic"
strange \cite{Bro96} and charm \cite{Bro81,Har96} of the proton.
We also need to check the applicability
for the statistical method to model the structure of the nucleon.
The total number of intrinsic partons inside the proton is around
\begin{eqnarray}
N&=&u_{\rm val}+d_{\rm val}+u_{\rm sea}+\bar{u}+d_{\rm
sea}+\bar{d}+g\\&=&2+1+2\bar{u}+2\bar{d}+g=5.5,
\end{eqnarray}
which is a small number of particles for the feasibility of
the statistical method. The success
of this study suggests us to apply the statistical method to the hadronic structure studies,
though its range of applicability should be very limited. It is
necessary to apply the statistical method from more sophisticated
consideration. Similar application can be also extended to other baryons.

In summary, we studied the proton structure as an ensemble of
quark-gluon Fock states. Using the principle of detailed balance,
the probabilities of finding the proton in every Fock states are
obtained without any parameter. A new origin of the light flavor
sea quark asymmetry, i.e., $\bar{u} \not= \bar{d}$, is given as a
pure statistical effect. It is found that $\bar{d}-\bar{u} \approx
0.124$, which is in surprisingly agreement with the experimental
observation.

We are very grateful to Prof.~Li-Ming Yang for his supervision and
encouragement, and to Dr.~Wei-Zhen Deng for helpful discussion.
This work is partially supported by National Natural Science
Foundation of China.

\mediumtext
\begin{table}
  \caption{The probabilities, $\rho_{i,j,k}$,
  of finding the quark-gluon Fock states of the
  proton, calculated using the principle of detailed balance without any parameter.
   $|\{q\},\{i,j,k\}\rangle$
   is the subsemble of Fock states, $i$ is the number of $u\bar{u}$ quark pairs,
   $j$ is the number
   of $d\bar{d}$ pairs, and $k$ is the number of gluons. }
        \label{table1}
        \begin{tabular}{ccccccccc}
        {i}&{j}&{$|\{q\},\{i,j,0\}\rangle$}&{$\rho_{i,j,0}$}&{$\rho_{i,j,1}$}&{$\rho_{i,j,2}$}&{$\rho_{i,j,3}$}
        &{$\rho_{i,j,4}$}&{$\cdots$}\\
    \tableline
        0&0&$|uud\rangle$         & 0.167849&0.167849&0.083924&0.027975&0.006994&$\cdots$\\
        1&0&$|uud\bar{u}u\rangle$ &0.055950&0.055950&0.027975&0.009325&0.002331&$\cdots$\\
        0&1&$|uud\bar{d}d\rangle$ &0.083924&0.083924&0.041962&0.013987&0.003497&$\cdots$\\
        1&1&$|uud\bar{u}u\bar{d}d\rangle$ &0.027975&0.027975&0.013987&0.004662&0.001166&$\cdots$\\
        0&2&$|uud\bar{d}d\bar{d}d\rangle$ &0.013987&0.013987&0.006994&0.002331&0.000583&$\cdots$\\
        2&0&$|uud\bar{u}u\bar{u}u\rangle$ &0.006994&0.006994&0.003497&0.001166&0.000291&$\cdots$\\
        1&2&$|uud\bar{u}u\bar{d}d\bar{d}d\rangle$ &0.004662&0.004662&0.002331&0.000777&0.000194&$\cdots$\\
        2&1&$|uud\bar{u}u\bar{u}u\bar{d}d\rangle$ &0.003497&0.003497&0.001748&0.000583&0.000146&$\cdots$\\
        0&3&$|uud\bar{d}d\bar{d}d\bar{d}d\rangle$ &0.001166&0.001166&0.000583&0.000194&0.000049&$\cdots$\\
        3&0&$|uud\bar{u}u\bar{u}u\bar{u}u\rangle$ &0.000466&0.000466&0.000233&0.000078&0.000019&$\cdots$\\
         $\cdots$&$\cdots$&$\cdots$&$\cdots$&$\cdots$&$\cdots$&$\cdots$&$\cdots$&$\cdots$\\
         \end{tabular}
   \end{table}

\mediumtext
\begin{table}
  \caption{The probabilities of finding the quark-gluon Fock states of the
  proton, with the process $g\Leftrightarrow gg$ considered.}
        \label{table2}
        \begin{tabular}{ccccccccc}
        {i}&{j}&{$|\{q\},\{i,j,0\}\rangle$}&{$\rho_{i,j,0}$}&{$\rho_{i,j,1}$}&{$\rho_{i,j,2}$}&{$\rho_{i,j,3}$}
        &{$\rho_{i,j,4}$}&{$\cdots$}\\
    \tableline
        0&0&$|uud\rangle$         &0.158885&0.158885&0.090791&0.037830&0.012610&$\cdots$\\
        1&0&$|uud\bar{u}u\rangle$ &0.052962&0.052962&0.028888&0.011234&0.003457&$\cdots$\\
        0&1&$|uud\bar{d}d\rangle$ &0.079442&0.079442&0.043332&0.016851&0.005185&$\cdots$\\
        1&1&$|uud\bar{u}u\bar{d}d\rangle$ &0.026481&0.026481&0.014123&0.005296&0.001558&$\cdots$\\
        0&2&$|uud\bar{d}d\bar{d}d\rangle$ &0.013240&0.013240&0.007062&0.002648&0.000779&$\cdots$\\
        2&0&$|uud\bar{u}u\bar{u}u\rangle$ &0.006620&0.006620&0.003531&0.001324&0.000389&$\cdots$\\
        1&2&$|uud\bar{u}u\bar{d}d\bar{d}d\rangle$ &0.004413&0.004413&0.002323&0.000852&0.000243&$\cdots$\\
        2&1&$|uud\bar{u}u\bar{u}u\bar{d}d\rangle$ &0.003310&0.003310&0.001742&0.000639&0.000183&$\cdots$\\
        0&3&$|uud\bar{d}d\bar{d}d\bar{d}d\rangle$ &0.001103&0.001103&0.000581&0.000213&0.000061&$\cdots$\\
        3&0&$|uud\bar{u}u\bar{u}u\bar{u}u\rangle$ &0.000441&0.000441&0.000232&0.000085&0.000024&$\cdots$\\
         $\cdots$&$\cdots$&$\cdots$&$\cdots$&$\cdots$&$\cdots$&$\cdots$&$\cdots$&$\cdots$\\
         \end{tabular}
   \end{table}

\end{document}